\documentclass[aps,prd,twocolumn,10pt,dlspace]{revtex4}
\usepackage{color}
\usepackage{epsfig}
\usepackage{graphicx}
\newcommand{\be}{\begin{equation}}
\newcommand{\ee}{\end{equation}}
\newcommand{\ba}{\begin{eqnarray}}
\newcommand{\ea}{\end{eqnarray}}
\newcommand{\nn}{\nonumber\\}
\begin{document}
\begin{flushleft}
TIFR/TH/10-17
\end{flushleft}
\title{Quarkonia in anisotropic hot QCD medium in a quasi-particle model}
\author{Vinod Chandra$^{a}$}
\email{vinodc@theory.tifr.res.in}
\author{V. Ravishankar$^{b}$}
\email{vravi@iitk.ac.in}
\affiliation{$^{a}$ Department of Theoretical Physics, Tata Institute of Fundamental Research,
Homi Bhabha Road, Mumbai-400005, India.}
\affiliation{$^b$ Department of Physics, Indian Institute of Technology Kanpur, Kanpur-208016, India.}
\date{\today}
\begin{abstract}
The fate of heavy quarkonia states has been investigated in  QCD at high temperature
when the plasma has a small momentum space anisotropy  within a quasi-particle model. 
A real time static potential has been obtained from a
Hard thermal loop expression for the gluon self-energy, by employing the quasi-parton 
equilibrium distribution functions extracted from hot QCD equations of state of  $O(g^5)$ and $O(g^6\ln(1/g))$.
Employing the potential, quarkonia melting has been studied. It is found that the equations of state and the momentum anisotropy cause significant modifications in the screening properties of the plasma.

\vspace{3mm}
\end{abstract}
\maketitle

\section{Introduction}
The dissociation of heavy quarkonia states due to the presence of screening of 
static color fields in hot QCD medium has long been 
proposed as a signature of a deconfined medium, and QGP formation\cite{matsui}.
Since than, this has been a area of active research\cite{nora} and undergone several refinements\cite{mocsy,umeda,patra,patra1,sdatta,nora1}. However, a precise understanding of this phenomenon is still
a subject of debate. There are 
disagreements between various approaches due to several ambiguities\cite{alberico}.
Attempts are being made on to fix them\cite{umeda}. A precise definition of the 
dissociation temperature is still elusive and is a matter of intense
theoretical and phenomenological 
investigations --from the perspective of lattice spectral function studies\cite{sdatta,spectral,petre,spt1,spt2} or potential inspired models\cite{alberico,pot1,pot2,mocsy1} or effective field theories\cite{eff}.
Experimentally, quarkonia suppression has been seen in RHIC experiments at BNL\cite{phenix} and hopefully the understanding will be refined at LHC heavy ion program in CERN in the near future. 

The heavy quark-antiquarks, such as $c\bar{c}$ are bound together by almost static gluons\cite{laine,dum,laine1}.
The fate of $q\bar{q}$ bound states, in  a QGP medium, can be explored by 
determining the behavior of gluon self energy at high temperature.
Recall that the real part leads to Debye screening on which the  traditional understanding of 
quarkonia melting in QGP medium is based.
The imaginary part of the potential
implies that thermal effects generate a finite width
for the quarkonium peak in the dilepton production rate.
They\cite{laine1} further showed that the physics related to the
finite width originates from the Landau-damping of low-frequency gauge fields, and could be
studied non-perturbatively by making use of the classical approximation.
This leads to a new proposal to explain the quarkonia melting in 
QGP medium\cite{laine1,laine2}. According to which the quarkonia bound states dissociate effectively 
at lower temperatures as compared to that obtained from Debye screening picture.
In this picture, the quakonium state will disappear whenever the thermal width
exceeds the binding energy. The temperature lies in the range $g^2M<T<gM$, where 
$M$ is the heavy quark pole mass\cite{laine}. The estimate has been refined 
employing the understanding from the effective theories in \cite{mas,laine_nucl,wu}.
These studies suggest that the quarkonia dissociation in the thermal medium is not dominated by the 
Debye screening rather by the above mentioned mechanism. 
Further, modifications to these estimates by considering the anisotropic momentum distribution 
for hard partonic degrees of freedom has been reported in Ref.\cite{laine}.

At this juncture, we note that the estimates of \cite{laine} for the dissociation temperature 
in the above analyses is based on the assumption that the QGP medium is fully equilibrated and that 
it can be treated as an ideal gas of quarks and gluons.
On the other hand, one knows that the latter assumption is particularly far from being true.
We are also in possession of equations of state for finite temperature QCD which have 
been evaluated up to $O(g^5)$\cite{zhai,arnold} and $O(g^6\ln(1/g))$\cite{kajantie}.
It is, therefore, natural that we should revisit the problem where
the strongly interacting nature of the quarks and gluons
in their deconfined state is explicitly and fully incorporated
in determining this important quantity. Taking
into consideration that the QGP which is produced in
heavy ion collisions does not possess the full isotropy, but
only a cylindrical symmetry at most, it is also not out of
place to study screening phenomenon in an anisotropic QGP medium.
This sets the motivation for our investigations.

In the present paper, we shall consider an anisotropic QGP (non-equilibrium environment)
and employ a recently proposed quasi-particle model for hot QCD equations of state\cite{chandra1,chandra2}.

We argue that all the assumptions taken in Ref.\cite{laine} will be equally valid in the present case 
too. Here, the anisotropy is in the momentum distribution functions of the hard partonic
degrees of freedom same as in Ref.\cite{laine}. This situation may be expected in the initial 
stages of the collisions due to the Bjorken expansion and extensively studied. 
There is a wealth of literature present\cite{roma,stric,stric1,stric2,bair,guo} where the impact of the anisotropy in various observables for QGP has been well studied. In most of these studies, the isotropic distribution function 
have been considered as the combination of the ideal Bose/Fermi distributions\cite{rebhan}.
The assumption that the hard degrees possess anisotropic momentum distribution
functions, is only justifiable at the late stages of the collision, so that soft degrees of freedom
have already get time to thermalize. As correctly stated in Ref.\cite{laine}, this does not
lead to a self-consistent non-equilibrium ground state and the system has techyonic modes.
Nevertheless, as emphasized in Ref.\cite{laine}, these issues will be absent in the small anisotropy limit.
Therefore, in view of this, we only restricts in the small anisotropy limit.

Here, we shall consider the more realistic distribution functions which are 
extracted from the improved perturbative QCD equations of state\cite{chandra1,chandra2}.
In this manner, we include the interactions in the various observables for QGP in the
presence of anisotropy. More precisely, we  address questions such as, (i) the form of the heavy quark potential with the realistic equations of state, and (ii) the dependence of the dissociation temperature of quarkonia states
on the anisotropy and interactions. We shall see that the interactions non-trivially modify the 
static heavy quark potential and the dissociation temperatures.
 
The paper is organized as follows. In section II, we shall review 
a quasi-particle description of hot QCD EOS which will be employed in the 
subsequent sections to study the fate of quarkonia in an anisotropic QGP. In section III, we 
shall discuss the heavy quark potential and melting of heavy quarkonia states. In section IV, we shall conclude the
present work.

\section{Quasi-particle description of hot QCD}
In this section, we briefly review a quasi-particle description of 
hot QCD equations of state proposed in Refs.\cite{chandra1,chandra2}. In these studies, a quasi-particle
model has been developed to capture all the interaction effects present in hot QCD equations of state in terms
of non-interacting quasi-gluons and quasi-quarks having temperature dependent 
effective fugacities. The model came out to be quite successful to yield the hot QCD equations 
of state accurately. The model has further been employed to 
study the chromo-electric response functions\cite{chandra2} and transport properties 
of QGP in RHIC\cite{chandra3,chandra4}. In this context, pure $SU(3)$ lattice EOS has also been 
studied\cite{chandra4}. A very strong prediction of the model is regarding the 
transport properties of QGP. In particular, the model has been employed to study the shear viscosity, $\eta$ of QGP\cite{chandra3,chandra4}. We obtained $\eta$ by employing the model in an appropriate transport equation\cite{bmuller,chandra3}. We showed that both  $\eta$ and its ratio with entropy density, $\eta/s$ are very sensitive 
to the interactions and they could be thought of as good diagnostics to 
distinguish various EOS at RHIC and LHC. A very small value for the ratio $\eta/s$
was obtained. We found that $\eta/s$ may violate the universal bound 
conjectured from AdS/CFT studies\cite{dtson}. 
 
Here, we consider the quasi-particle description\cite{chandra1,chandra2} of $O(g^5)$ 
hot QCD\cite{zhai,arnold}  and $O(g^6\ln(1/g)$ hot QCD EOS \cite{kajantie}.
We denote former EOS as EOS1 and latter as EOS2. EOS1 and EOS2 
qualitatively matches with the more realistic lattice EOS for $T\ge 2\ T_c$.
In fact, EOS2 shows better agreement as compared to EOS1\cite{chandra2}.
Note that EOS1 and EOS2 have been evaluated at 
vanishing baryon density. 
We start with the equilibrium distribution functions extracted from these two equations of state.
They posses a very simple form,
\ba
\label{eq1}
f^{g}_0=\frac{z_g\exp(-\beta p)}{\bigg(1-z_g\exp(-\beta p)\bigg)}\nn
f^{q}_0=\frac{z_q\exp(-\beta p)}{\bigg(1+z_q\exp(-\beta p)\bigg)},
\ea
where $g$ stands for quasi-gluons, and $q$ stands for quasi-quarks. $z_g$ is the quasi-gluon effective fugacity and 
$z_q$ is quasi-quark effective fugacity. These distribution functions are isotropic in nature. 
At this juncture, we want to emphasize the physical meaning of  $z_g$ and $z_q$.
These fugacities should not be confused with any conservations law (number conservation) and have merely been introduced to encode all the interaction effects at high temperature QCD. 
At this juncture, we wish to emphasize that the effective fugacities with the similar motivations 
to address complicated interactions in BEC (Bose Einstein Condensate)\cite{bose} and
interacting Fermi systems\cite{fermi} have already been 
introduced in condensed matter physics. 

 Their physical meaning could be understood in terms of modified quasi-parton dispersion relations,
\ba
\label{disp}
\omega^g_p=p+T^2\partial_T \ln(z_g)\nn
\omega^q_p=p+T^2\partial_T \ln(z_q),
\ea
where $\omega_p$ denotes the single quasi-parton energy. It is clear from Eq. (\ref{disp}) that 
$z_{g/q}$ non-trivially contribute to the single particle energy. This tells us that the model is
in the spirit of Landau theory of Fermi liquids. Further, $z_{g/q}$ leads to the quasi-parton number densities
$N_g$ and $N_q$ as follows,
\be
N_{g/q}=\int \frac{d^3p}{8\pi^3} f^{g/q}_0=\frac{\pm \nu_{g/q}}{\pi^2} PolyLog[3,\pm z_{g/q}],
\ee
where the function $PolyLog[3,z]=\sum_{l=1}^\infty \frac{z^l}{l^3}$. Both $z_g$ and $z_q$ have a very complicated 
temperature dependence and asymptotically reach to the ideal value unity\cite{chandra2}.
Here, $\nu_g\equiv 2(N^2_c-1)$ and $\nu_q\equiv 4 N_c N_f$, are the number of gluonic and
quark-antiquark degrees of freedom respectively for $SU(N_c)$.
The quasi-particle description does not change the group velocity of quasi-partons, 
$\vec{v}^{g/q}_{p}= \partial_{\vec{p}} \omega^{g/q}_p\equiv \hat{p}$ (this observation may play an
important role in setting up the kinetic equation to determine the transport parameters for QGP with the 
quasi-particle model). This property allows us to distinguish the present model from other models based on 
an effective mass description\cite{peshier,pesh1}.

This description, which enables us to adapt realistic QCD EOS to make predictions for RHIC, leads
to very very interesting results as far as the bulk and transport properties of 
QGP in RHIC concern.
Encouraged by that we now seek to employ the understanding to study the fate of quarkonia in a anisotropic QGP medium. 

\section{Heavy quarkonia in anisotropic QGP}
It is phenomenologically quite interesting to study
QGP system which is slightly away from equilibrium.
It is interesting to inspect quarkonia in a hot QCD environment where the
hard parton distribution functions posses anisotropy in their momentum distribution function.
This phenomenon is inspected in the recent past in  Ref.\cite{laine} and very interesting physical 
understanding has been obtained. Further, this problem has also been considered for an ideal EOS
in Ref.\cite{dum}.These results deviate slightly from \cite{laine} because of the fact that the 
physics setting is different. They\cite{dum} also assumed non-thermal distributions for the soft glouns.
For similarities  between these two analyses, we refer the reader to \cite{laine}.

Here, we shall primarily follow the work of \cite{laine}  and extend in the 
case of EOS1 and EOS2 by combining the analysis with the quasi-particle description discussed
in the previous section. The main motivation here is to 
adapt the realistic hot QCD EOS and investigate the fate of quarkonia bound states
in this interacting  environment.

\subsection{The real time static potential}
Here, we combine the quasi-particle description of hot QCD
described in the previous section to the formalism of heavy quark-potential in 
anisotropic QGP\cite{laine}. The aim here is to see how the inclusion of interactions
coming from the realistic equations of state for QGP affect  the 
heavy quark-potential and quarkonia melting in an anisotropic QGP.

The starting point to obtain the real time static quark-antiquark potential is the Hard thermal loop gluon self energy, $\Pi^{\mu,\nu}$ for a general momentum distribution function\cite{rebhan,thoma}, $f(\vec{p})$,

\be
 \Pi^{\mu\nu}_R(K) = g^2 
 \int \! \frac{{\rm d}^3\vec{p}}{(2\pi)^3} \,
 v^\mu \partial^\alpha f(\vec{p}) 
 \bigg( {\delta^\nu}_\alpha - \frac{v^\nu k_\alpha}{v^\sigma k_\sigma + i 0^+} \bigg) 
 \;
\ee
where $v\equiv (1,\vec{p}/p)$, and  $R$ indicates that the self-energy appears in the inverse
of the retarded propagator. The hard mode momentum distribution has been assumed to be the following form\cite{prd_aniso}, 
\be
\label{eqan1}
f(p) = f_{0}(\sqrt{p^2 + \xi(\vec{p} \cdot\hat{n})^2}),
\ee
where $\xi$ a real parameter known as the anisotropy parameter and $\hat{n}$ is the 
direction of anisotropy. Without the loss of generality, $\hat{n}$ can be  chosen to be along the z-direction. The quantity  $\vec{p}\cdot\hat{n}\equiv p_z$, where $p_z$ is the momentum component in the beam direction in RHIC.
The distribution functions $f (p)$ can be constructed from an arbitrary isotropic distribution
function, $f_{0}$ by the rescaling of only one direction in
momentum space. We note that $\xi$ is in $[-1,\ \infty]$,
and $\xi=0$ corresponds to the 
isotropic case. The  values corresponding to the Bjorken expansion induced anisotropy are $\xi > 0$.

Now, making use of the change of variable
$\bar{p} \equiv \sqrt{p^2 + \xi p_z^2}$, and 
denoting $p_z \equiv p \cos\theta_p$, and following the analysis in Ref.\cite{laine}, 
we get the following form of the component $\Pi^{00}_R$ (which is needed for the 
real time static potential) near the static limit assuming the 
anisotropy, $\xi$ to be small (first order in $\xi$)\cite{laine},  
\ba
\label{pi0}
 \Pi^{00}_R(K) 
 &&\approx m_D^2 \bigg\lbrace 
 -1 + \xi \bigg[ \frac{1}{6} - \frac{1}{2} \cos (2\theta_k) \bigg]
\nn&&
 - \frac{i \pi}{2} \frac{k_0}{k}
 \biggl[ 
 1 + \xi   \cos (2\theta_k)
 \biggr]
 \bigg \rbrace 
,
\ea
where $\theta_k$ is the angle between $\vec{k}$ and $\vec{z}$. The quantity $m_D$ is defined in terms of the 
equilibrium (isotropic) distribution function as, 
\be
\label{debye}
 m_D^2 \equiv -g^2 
 \int  \frac{{\rm d}^3\bar{\vec{p}}}{(2\pi)^3} \, 
 \frac{{\rm d}f_{iso}(\bar{p})}{{\rm d}\bar{p}}.
\ee
So far the analysis is quite general and allow any suitable arbitrary form for the 
$f_{iso}$.  Note that often, $f_{iso}$ is taken to be a 
combination of ideal Bose-Einstein and Fermi-Dirac distribution functions as\cite{rebhan},
\be
\label{eq8}
f_{iso}=2 N_c n_{b}(p) + 2 N_f (n_q(p) + n_{\bar{q}}(p)),
\ee
 where $n_b(p)=\frac{1}{exp(-\beta p)-1)}$, and in the zero baryon chemical potential 
$n_{q}=n_{\bar{q}}=\frac{1}{exp(-\beta p)+1)}$. This combination leads to the leading order HTL expression 
($m_D^2=g^2(T) T^2 (N_c/3 +N_f/6)$) for the Debye mass in hot QCD. Here, $N_c$ denotes the number of colors and
$N_f$ the number of flavors.

A more appropriate choice for $f_{iso}$ would be a suitable combination of quasi-parton distribution functions\cite{chandra2} ($f^{g}_0$ and $f^q_{0}$ displayed in Eq. (\ref{eq1}).). These are mandated by the fact that QGP is not an ideal gas of gluons and quarks rather a strongly interacting system\cite{star,boyd,karsch,cheng,fodor,fodor1}. 

We make the following substitutions for $f_{iso}$,
\ba
\label{iso}
f_{iso}=
%\bigg\lbrace 2 N_c f^g_{0}; \ 2 N_c f^{g}_0+2 N_f f^{q}_0\bigg\rbrace.
\left\{ \begin{array}{rcl}
2 N_c f^g_{0} &\mbox{for pure gauge,} &\\
2 N_c f^{g}_0+2 N_f f^{q}_0&\mbox{for full QCD}&
\end{array} \right.
\ea
Note that the expression for the Debye mass in Eq.(\ref{debye}) with $f_{iso}$ in Eq.(\ref{eq8}) is same as leading order 
HTL expression  obtained from effective kinetic models  \cite{blaizot,blaizot1,kelly,kelly1} for hot QCD medium
with ideal EOS. These results are equally applicable in the present case too due the fact the
the quasi-partons in our quasi-particle model are non-interacting. The only consistency requirement is that 
in the asymptotic limit the results should match with those for the ideal EOS. 
This could be achieved with the chosen expressions for $f_{iso}$ in Eq.(\ref{iso}).
We shall discuss the Debye mass with these choices of $f_{iso}$ in detail in the next subsection.

We now discuss the real and imaginary part of the static potential which depends upon the 
form of $m_D$ and the anisotropy parameter. In the leading log order 
the real and imaginary parts have been analyzed in Ref.\cite{laine}, and they
obtain the following form,
\ba
\label{pot}
Re[V>]&&\sim -\frac{g^2 C_F}{4\pi r}[1+O(\hat{r})]\vert_{r\sim \frac{1}{g^2 M}}\nn
Im[V>]&&\sim -g^2 C_F T\frac{m_D^2 r^2}{3}\ln(\frac{1}{m_D r})[1+O(\frac{1}{ln(\hat{r})})]
\nn&&\times (1-\frac{\xi}{3})
\ea
where M is the pole mass of heavy quarks\cite{laine}, and $\hat{r}= m_D r$. Here, only the s-wave ground states 
have considered. In this case the term proportional to $(1-3 cos^2 \theta_k)$ in Eq.(\ref{pi0}) average out to zero\cite{laine} at first order in perturbation theory and corrections will be of the order $\xi^2$.
Similar mathematical forms for the real and imaginary part of the static potential 
are obtained for the present case too. The interactions appear via $m_D$ and 
merely renormalizes the effective charges of the quasi-partons, which we shall see next.
   
\subsubsection{The Debye mass $m_D$}
Before, we employ the quasi-parton distribution functions in Eq.(\ref{eq1}) to 
obtain the temperature dependence of $m_D$, We wish to discuss a very important point here. There is an issue of defining the parameter $\beta\equiv 1/T^\prime$ which appears in $f_{iso}$ in terms of the equilibrium temperature $T$\cite{laine}.
The parameter $T^\prime$ will be different  from the temperature $T$ in the equilibrium. 
This is due to the fact that, in the presence of anisotropy the system is away from equilibrium. 
If we assume the system to be very close to the equilibrium (case of very small anisotropy) then the parameter $T^\prime$
could be taken to be $T$. As discussed in \cite{laine}, the parameter $T^\prime$ can also be fixed 
from isentropic condition. This is quite tedious here due to the non-trivial temperature dependence of the effective fugacities, $z_g$ and $z_q$. We do not consider this 
particular case here and stick to the near equilibrium condition.

Now, employing $f^g_{0}$ and $f^q_{0}$ in Eq.(\ref{debye}) we obtain,
$m_D$ in the purely gluonic case and full QCD as,
\ba
\label{dby}
m^2_D &=& g^2 T^2 \bigg( \frac{N_c}{3}\times \frac{6 PolyLog[2,z_g]}{\pi^2}\bigg) \mbox{for pure gauge}\nn
m^2_D &=& g^2 T^2 \bigg\lbrace\bigg( \frac{N_c}{3}\times \frac{6 PolyLog[2,z_g]}{\pi^2}\bigg)\nn
&&+\bigg(\frac{N_f}{6}\times\frac{-12 PolyLog[2,-z_q]}{\pi^2}\bigg)\bigg\rbrace \mbox{for full QCD}. \nn
\ea 
Here, $g(T)$ is the QCD running coupling constant, $N_c=3$ ($SU(3)$) and $N_f$ is the number of flavor, the function
$PolyLog[2,z]$ having  form, $PolyLog[2,z]=\sum_{k=1}^{\infty} \frac{z^k}{k^2}$.
These are the same expression which we have obtained from the chromo-electric response functions\cite{chandra2}
for the interacting QGP. The extra factors appearing due to the inclusion of interactions can be attributed to the charge renormalization in QGP medium\cite{chandra2}. Define the effective charges as,
\ba
 Q^2_g&=& g^2 \frac{6 PolyLog[2,z_g]}{\pi^2}\nn
 Q^2_q&=& g^2 \frac{-12 PolyLog[2,-z_q]}{\pi^2}.
\ea
\begin{figure*}
\vspace{5mm}
\includegraphics[scale=.40]{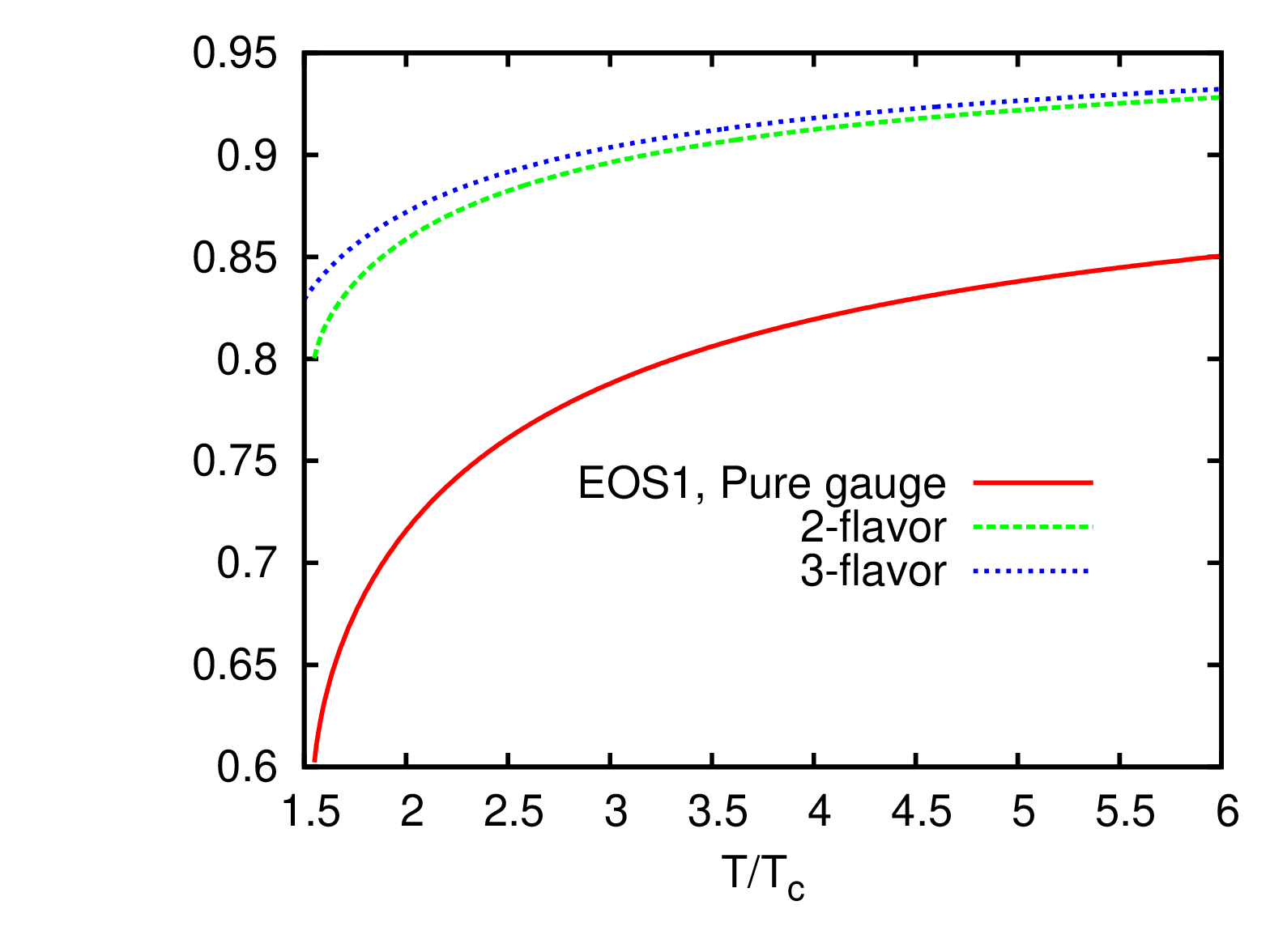} 
\hspace{-.45cm}
\includegraphics[scale=.40]{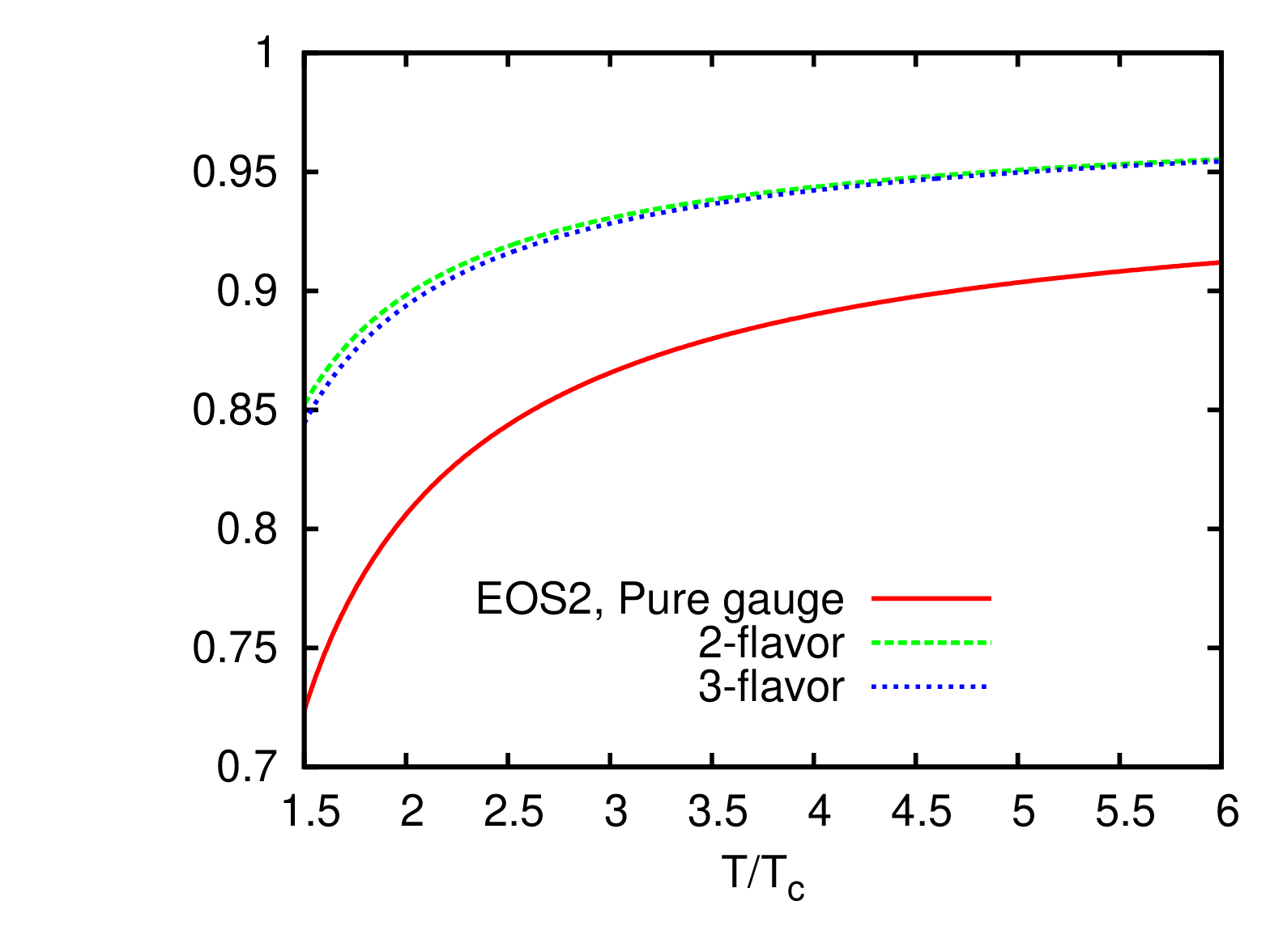}
\hspace{2mm}
\caption{Behavior of $\mu_d$ as function of $T/T_c$. The left panel
shows the results for EOS1 and right panel for EOS2. In both the cases, $\mu_d$ reaches the ideal value unity only 
asymptotically, and $\mu_d\le 1$.}
\vspace{5mm}
\end{figure*}
The expressions for the Debye mass 
can be recasted as,
\ba
\label{eff}
m^2_D=
%\bigg\lbrace Q^2_g T^2\frac{N_c}{3}; \ T^2( \frac{N_c}{3} Q^2_g+\frac{N_f}{6} Q^2_q)\bigg\rbrace.
\left\{ \begin{array}{rcl}
Q^2_g T^2\frac{N_c}{3} &\mbox{for pure gauge,} &\\
T^2( \frac{N_c}{3} Q^2_g+\frac{N_f}{6} Q^2_q) &\mbox{for full QCD}&
\end{array} \right.
\ea
Both $Q^2_g$ and $Q^2_q$ acquire the ideal value $g^2$ only asymptotically. Moreover, $\lbrace Q^2_g, Q^2_q\rbrace \le g^2(T)$.  In other words, these expressions for the Debye masses reach their ideal values (leading order HTL expression) only asymptotically). To see this, we plot the ratio $\mu_d\equiv m_D/m^{I}_D$,
\ba
\mu_d=
%\bigg\lbrace \frac{m_D}{g(T) T};\ \frac{m_D}{g(T) T(1+\frac{N_f}{6})}\bigg\rbrace \nn
% \equiv\bigg\lbrace \frac{Q_g}{g}; \frac{(\frac{Q_g}{g}+\frac{N_f}{6}\frac{Q_q}{g})}{(1+\frac{N_f}{6})}\bigg\rbrace,
\left\{ \begin{array}{rcl}
 \frac{Q_g}{g}&\mbox{for pure gauge,} &\\
\frac{(\frac{Q_g}{g}+\frac{N_f}{6}\frac{Q_q}{g})}{(1+\frac{N_f}{6})} &\mbox{for full QCD}&
\end{array} \right.
\ea
as a function of temperature employing the temperature dependence of $z_g$ and $z_q$ from Ref.(\cite{chandra2})
in Eq.(\ref{eff}) for pure gauge theory and full QCD respectively in Fig. 1.

The quantity $\mu_d$ scales with $T/T_c$ coming from the scaling properties of 
$z_g$ and $z_q$\cite{chandra2}.
The ratio, $\mu_d$ is relevant in investigating the dissociation temperatures for 
quarkonia states, which we shall see next.
\begin{figure*}
\vspace{5mm}
\includegraphics[scale=.40]{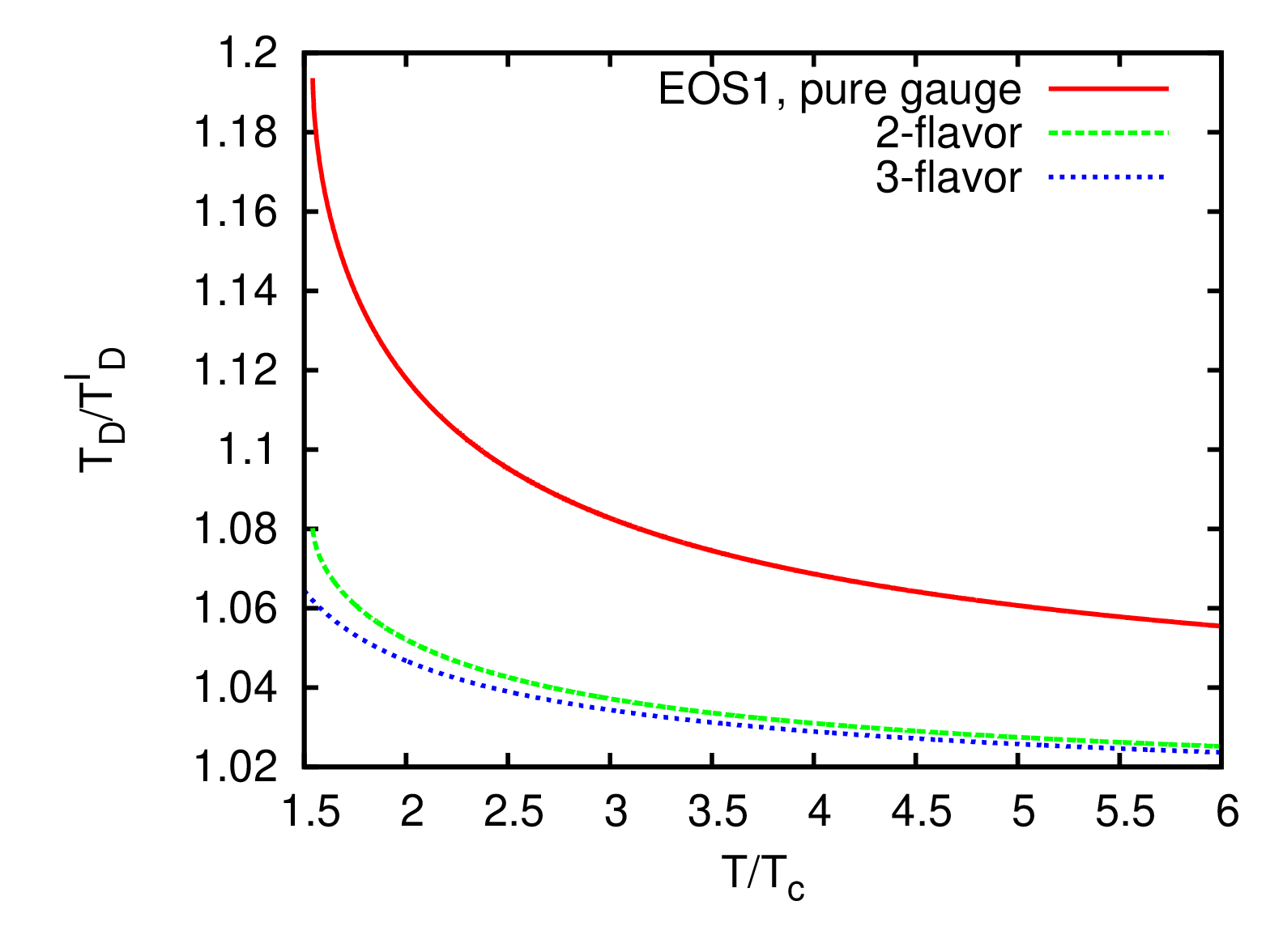} 
\hspace{-.45cm}
\includegraphics[scale=.40]{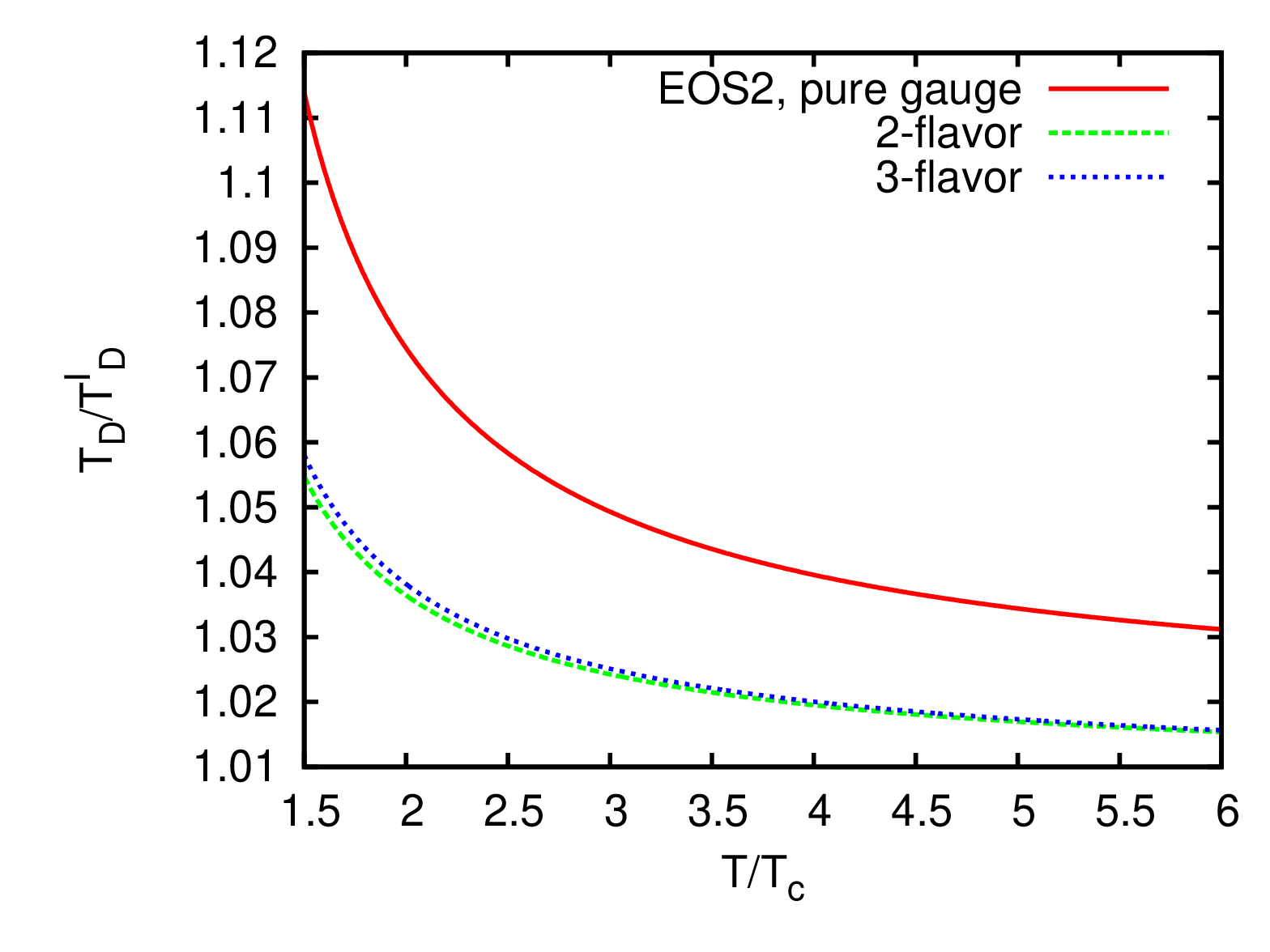}
\hspace{2mm}
\caption{Behavior of $T_D/T^I_D$ as function of $T/T_c$. The left panel
shows the results for EOS1 and right panel for EOS2. In both the cases, it reaches the ideal value unity only 
asymptotically, and always $\ge 1$.}
\vspace{5mm}
\end{figure*}

\subsection{Quarkonia dissociation}
Here, we choose the following criteria for the quarkonia dissociation.
The temperature at which the  magnitude of the real part of the static potential (real part gives the binding energy) displayed in Eq.(\ref{pot}), equals the magnitude of the  imaginary part of the potential (imaginary part gives the width of the state) is identified as the dissociation temperature. 

Employing Eq.(\ref{pot}), We obtain the following condition, 
\be
\frac{g^2}{r}\sim g^2 \frac{T m_D^2 r^2}{3}\ln(\frac{1}{m_D r})\times(1-\frac{\xi}{3}).
\ee
Where $r\sim 1/{g^2 M}$\cite{laine}. Using $T^\prime= T$ and Debye masses given in Eq.(\ref{dby}), we obtain the 
following expression for the melting temperature in pure gauge theory,
\be
\label{tdg}
T_{D}=\frac{g^{4/3} M}{3^{1/3}(\frac{Q_g}{g})^{\frac{1}{3}}}\times \bigg(\frac{1}{2}\ln(\frac{g}{Q_g})+\ln(\frac{1}{g})\bigg)^{-\frac{1}{3}}\times(1+\frac{\xi}{9}).
\ee
On the other hand in full QCD,
\ba
\label{tdqg}
T_{D}&=&\frac{g^{4/3} M}{3^{1/3}(\frac{Q_g}{g}+\frac{N_f}{6}\frac{Q_q}{g})^{\frac{1}{3}}} \bigg(\frac{1}{2}\ln(\frac{1}{(\frac{Q_g}{g}+ \frac{N_f}{6} \frac{Q_q}{g})})+\ln(\frac{1}{g})\bigg)^{-\frac{1}{3}} \nn&&\times(1+\frac{\xi}{9}).
\ea

The corresponding ideal expression for pure gauge theory reads,

\be
T^{I}_{D}=\frac{g^{4/3} M}{3^{1/3}}\times \bigg(\ln(\frac{1}{g})\bigg)^{-\frac{1}{3}}\times(1+\frac{\xi}{9}).
\ee

On the other hand in full QCD,
\ba
\label{td1}
T^{I}_{D}&=&\frac{g^{4/3} M}{3^{1/3}(1+N_f/6))^{\frac{1}{3}}}\times \bigg(\frac{1}{2}\ln(\frac{1}{(1+ N_f/6)})+\ln(\frac{1}{g})\bigg)^{-\frac{1}{3}}\nn&&\times(1+\frac{\xi}{9}).
\ea

Note that Eqs.(\ref{tdg}),(\ref{tdqg}) posses same $\xi$ dependence as in\cite{laine}. There the role played 
by $\xi$ in modifying the potential as well as the quarkonia melting temperatures will be exactly the same.
The interesting observations are regarding the 
inclusion of the realistic EOS (EOS1 and EOS2). As it is clear from Eqs.(\ref{tdg}) and (\ref{tdqg}) that interactions non-trivially 
modify the s-wave ground states quarkonia melting temperature. To see the effects of interactions 
more closely and quantitatively, we consider the ratio $T_D/T^I_D$ in the case of both pure gauge theory 
and full QCD. It is straight forward to see that in the weak coupling limit ($g<<1$), the ratio can be given 
as,
\ba
\label{td2}
\frac{T_D}{T^I_D}= 
%\bigg\lbrace (\frac{Q_g}{g})^{1/3};\
%\bigg(\frac{(\frac{Q_g}{g}+ \frac{N_f}{6} \frac{Q_q}{g})}{1+ \frac{N_f}{6}}\bigg)^{1/3}\bigg\rbrace.
\left\{ \begin{array}{rcl}
(\frac{Q_g}{g})^{-1/3} &\mbox{for pure gauge,} &\\
\bigg(\frac{(\frac{Q_g}{g}+ \frac{N_f}{6} \frac{Q_q}{g})}{1+ \frac{N_f}{6}}\bigg)^{-1/3} &\mbox{for full QCD}&
\end{array} \right.
\ea

Note that the ratio $T_D/T^I_D\equiv \mu_d^{-\frac{1}{3}}$.
Clearly from the Fig.1, the ratio is always $\ge 1$. 
It reaches to unity only asymptotically. Therefore, the dissociation temperatures with non-ideal EOS 
will always be higher as compared to those for their ideal counter parts.
This implies that the presence of interactions decreases the effect of Debye screening
in hot QCD. We have plotted the ratio, $T_D/T^I_D$ in Fig. 2.
Quantitatively, dissociation temperatures are
$\sim$ 20\%-6\% higher than the ideal value near $T\sim 1.5 T_c$ for EOS1.
On the other hand, for EOS2, these are $\sim$ 12\%-5\% higher than the ideal values near  $T\sim 1.5 T_c$ (see Fig. 2).
In the case of  2- and 3- flavor QCD, the dissociation temperatures are  more closer to their
ideal counter parts as compared to the pure gauge theory. This is true in the case of both EOS1 and 
EOS2. In other words, differences with the ideal EOS are more pronounced in the case of pure gauge theory.

In the anisotropic case with non-ideal EOS, the dissociation temperatures 
vary from the corresponding isotropic case by a factor $\frac{\xi}{9}$.
Therefore, in the case of non-ideal EOS (EOS1 and EOS2) for $\xi \sim 1$
there is an increase of  $\sim 10\%$ in the dissociation temperatures as compared to the equilibrium case.
These conclusions regarding the anisotropy are same as that for an ideal EOS in\cite{laine}.
This is not unexpected because the quasi-parton distribution functions employed in this work only differ
from the ideal Bose/Fermi distributions by purely temperature dependent 
suppression factors (effective fugacities, $z_g$, $z_q$) otherwise they are
non-interacting.

\section{Conclusions}
In conclusion, we have studied 
the form of the static inter-quark potential and 
quarkonia dissociation in an anisotropic QGP in the 
small anisotropy limit employing quasi-parton equilibrium distribution functions 
obtained from $O(g^5)$ and $O(g^6(\ln(1/g))$ hot QCD equations of state.
We find that the interactions non-trivially modify the real time static potential
for a heavy quark-antiquark pair. In particular, in the leading order 
the real part of the potential remains intact as for ideal EOS. In contrast,
imaginary part gets non-trivial modifications. Interestingly,
the dissociation temperatures are also non-trivially modified in the presence 
of interactions. We studied both pure gauge theory and full QCD sectors. 
We show that the interactions coming from the realistic QGP EOS 
merely renormalize the partonic charges. They enter in the potential as well as in 
the dissociation temperatures through the Debye mass.
The above conclusions are found to be valid for both the equations of states.
These results are valid for small anisotropies. For larger anisotropies, numerical simulations 
are required. It is clear from our study that for the non-equilibrium states at small momentum space anisotropy one can estimate the relative change caused by the anisotropy at leading-logarithmic
order employing the realistic QGP EOS. 

It would be of interest to extend the present study to the case of non-vanishing baryon
density and recently proposed realistic lattice equations of state with physical quarks masses\cite{cheng,fodor1},
and also the chromo-electric response functions\cite{chandra2}. 
Finally, it would be of interest to consider the Bjorken hydrodynamic expansion and
study the viscous corrections to the response functions. These investigations
will be done separately in near future.  A more interesting 
study would be to develop a two-component model which can handle anisotropies 
in both hard and soft partonic modes and their impact on the inter-quark forces in hot QCD 
medium.
 
\vspace{4mm}
\noindent{\bf Acknowledgement:} 
VC is thankful to Rajeev Bhalerao for many helpful discussions and 
Nidhi Gour for careful reading of the manuscript. He acknowledges Department of Physics, IIT Kanpur, India
for the hospitality under the TPSC initiative.

\end{document}